\documentclass[twocolumn,amsmath,amssymb,floatfix,nofootinbib]{revtex4}
\newcommand\be{\begin{equation}}
\newcommand\ee{\end{equation}}
\newcommand\ba{\begin{eqnarray}}
\newcommand\ea{\end{eqnarray}}
\newcommand\eq{\begin{equation}}           
\newcommand\en{\end{equation}}

\input epsf
\usepackage{amssymb}
   \usepackage{epsfig}
\begin{document}
\title{
{\hfill  \small FERMILAB-PUB-05-059-A \\ ~\\~\\}
Precision of Inflaton Potential Reconstruction from CMB\\  Using the General Slow-Roll Approximation}

\author{Kenji Kadota$^1$, Scott Dodelson$^{1,2,3}$, Wayne Hu$^{2,4}$ 
and Ewan D. Stewart$^{5,6}\footnote{On sabbatical leave from Department of Physics, KAIST, Republic of Korea.}$}
\affiliation{
$^1$ {\em  Particle Astrophysics Center, Fermi National Accelerator Laboratory, }
{\em Batavia, IL 60510, USA} }
\affiliation{
$^2${\em Department of Astronomy \& Astrophysics, }
{\em The University of Chicago, Chicago, IL 60637, USA}
}
\affiliation{
$^3${\em  Department of Physics and Astronomy,}
{\em Northwestern University, Evanston, IL 60208}
}
\affiliation{
$^4$ {\em Kavli Institute for Cosmological Physics,
The University of Chicago, Chicago, IL 60637, USA}
}
\affiliation{
$^5${\em Department of Physics, KAIST, Daejeon 305-701, Republic of Korea}
}
\affiliation{
$^6${\em Canadian Institute for Theoretical Astrophysics,
University of Toronto, Toronto, ON M5S 3H8, Canada}
}

\begin{abstract}

Through a principal component analysis, we study how accurately CMB observables can constrain inflaton potentials in a model independent manner. 
We apply the general slow-roll approximation in our analysis where we allow, in contrast to the standard slow-roll approximation, 
the possibility of variations in $V''(\phi)$ and take into account the fact that horizon crossing is not an instantaneous event. 
Our analysis provides a set of modes to be used
in fitting observables. We find that of order five of these modes will be constrained by 
future observations,
so a fully general data analysis package could use the amplitudes of just a handful of 
modes as free parameters
and retain all relevant information in the data. 
\end{abstract}

\maketitle

\setcounter{footnote}{0} 
\setcounter{page}{1}
\setcounter{section}{0} \setcounter{subsection}{0}
\setcounter{subsubsection}{0}

\section{Introduction}
The origin of cosmic density perturbations is one of the most 
notable predictions of the inflationary scenario. The observational 
data on their power spectrum from the cosmic microwave background (CMB) 
and large scale structure provide us with one of few windows through 
which we can probe the physics governing the early 
Universe \cite{wmap,york,max,selj}.
In particular it is interesting to ask if the CMB can directly 
constrain the potential energy of the field driving
inflation, the {\it inflaton}, and if so, 
which features of the potential will be most accurately measured.
Extracting the physics of inflation from CMB data 
alone is not trivial because different inflation models can produce 
identical CMB spectra and moreover CMB observables can cover only 
a tiny portion of an inflaton potential. 
However, we can still try to constrain the curvature, for example, of 
the inflaton potential during the time when cosmologically interesting 
scales were exiting the horizon, and the sign of the curvature alone can 
already distinguish between several classes of inflation models. 
Because we are still 
not certain of the functional form of the inflaton potential, it would 
be useful to work in a model-independent manner to avoid biasing our 
estimates of the properties of inflaton potentials. Further, we are
motivated to ask what other features of the potential are constrained
by the CMB, again in a model-independent way?

There exist many works which reconstruct the primordial power 
spectrum ${\cal P}(k)$ from the observable CMB data $C_l$ 
(see, for example, \cite{wayne2,CtoP1,CtoP2,CtoP3,CtoP4,CtoP5,CtoP6} for 
model-independent approaches).
If we can also measure the tensor perturbations and assume the standard slow-roll 
approximation, reconstructing the inflaton potential is feasible \cite{rocky}.
Even without the standard slow-roll approximation, the conversion from ${\cal P}(k)$ 
to inflaton potential parameters is possible analytically for a given 
${\cal P}(k)$ \cite{salman,joy}.
So we could combine these analyses to examine the properties of inflation 
from $C_l$ by two steps: $C_l \rightarrow {\cal P}(k) \rightarrow$ inflaton 
potential parameters. However, degeneracies between the parameters 
arise from these two steps, and it would be more efficient to constrain an 
inflaton potential directly from $C_l$. 
If we perform a numerical analysis such as the Markov Chain Monte Carlo method, 
reconstructing the inflaton potential directly from $C_l$ can be done without 
heavily relying on the standard slow-roll approximation \cite{liddle,wil}. 
However, even in the case of numerical analysis, how many inflaton potential 
parameters do we need in addition to cosmological parameters such as the reionization 
optical depth and baryon density? Certainly including more parameters would lead 
to a better fit to observations, but adding as many free parameters as we want 
would not be a practical approach. Our principal component analysis will be 
useful in choosing an optimal set of parameters to characterize
the inflaton potential.   

In Sec. \ref{sec1} we describe the general slow-roll approximation with 
an emphasis on the difference from the standard slow-roll approximation, 
and discuss a model-independent parametrization of the inflaton potential. 
Sec. \ref{principal} outlines a principal component analysis and determines 
the precision that idealized future CMB experiments can attain in constraining 
inflaton potential parameters. Sec.~\ref{examples} gives some illustrations
of the formalism, followed by the conclusion in Sec. \ref{conc}.
 
\section{Inflaton Potential Parametrization}
\label{sec1}
We, in this section, discuss a model independent parametrization 
of the inflaton potential (strictly speaking, a function of the inflaton potential) in view of the general slow-roll approximation to study how precisely CMB observables can ultimately constrain the inflaton potential.

\subsection{General Slow-Roll Approximation}
We refer the readers to the Appendix and Ref. \cite{ewangeneral} for the discussion of general slow-roll approximation, but let us briefly here point out the motivations for our using the general slow-roll approximation rather than the standard slow-roll approximation in our analysis.
 
One of the possible pitfalls of the standard slow-roll approximation is that it 
presumes the scale invariance of $V''(\phi)$ (in the rest of the paper, primes denote derivatives with respect to the argument). 
For single field inflation with slow-roll, the observations are indeed consistent with 
a small amplitude\footnote{We use units in which $8\pi G=1$.} of $V''$. Neither observations nor theories, however, require 
exact scale invariance. The general slow-roll approximation can lift this extra 
(and unnecessary) condition of the scale invariance of $V''$ \cite{ewangeneral}. This would help us consider the possibilities for a wider variety of inflaton potential properties from CMB observables. The application of the general slow-roll approximation in our principal component analysis where the standard slow-roll approximation is not applicable is given in the second example of Section \ref{examples} for the illustration purpose.  

Another pitfall of the standard slow-roll calculations is the matching condition at 
``horizon crossing'' $k={\cal O}(1)aH$ where one conventionally just evaluates the 
perturbations exactly at $k=aH$. The ${\cal O}(1)$ ambiguity in the ``horizon crossing'' \cite{pit} and also the possible prolonged effects around horizon crossing are accounted for in 
the general slow-roll approximation. Each Fourier mode in the perturbation calculations is affected by the inflationary dynamics over the whole
period it is leaving the horizon, not just at one instant of ``horizon crossing''. Therefore, the correlations of each Fourier mode can be highly non-trivial compared with the cases of the standard slow-roll. This makes our use of principal component analysis rather useful because it gives us the independent (i.e. statistically independent) modes for the clearer physical interpretation of the parameter estimations as discussed in Sec. \ref{principal}.

 \subsection{Discrete Parametrization}
\label{discpa}
Scalar perturbations alone cannot constrain $V$ itself 
(we need tensor perturbations to do it) but rather certain functions of $V$. In this paper, we consider the physical limitations 
of CMB observables without tensor perturbations for constraining\footnote{In this paper, we consider only single field inflation models. Our analysis can, however, be extended straightforwarly to the multiple field inflation models because the general slow-roll formula for multiple-component inflation has an analogous form as that for the single component case \cite{ewannew}.}  
\ba
G_n\equiv  3\left(\frac{V'}{V}\right)^2-2 \frac{V''}{V}~~~.
\ea
With the standard slow-roll approximation, the spectral index is directly related to this
quantity $G_n$ because $n-1= -3 (V'/V)^2+ 2  {V''}/{V}$. In our analysis, we use the general slow-roll approximation, in which case we do not have such a simple relation. $G_n$ is, however, still a fundamental quantity representing an inflaton potential even for the general slow-roll cases and indeed can be interpreted as the source function for $n-1$ (see Appendix for the derivation) 
\ba
n-1 &\equiv& \frac{d \ln {\cal P}}{d \ln k} \nonumber \\
&=& \int^{\xi_{\rm max}}_{\xi_{\rm min}} \frac{d \xi}{\xi} [-k\xi W'(k \xi)]\left[-G_n\right].
\ea
where $\xi$, minus the conformal time, is an integral over the scale factor $a$
up to the end of inflation:
\ba
\xi \equiv - \int_t^{t_{\rm end}} {dt'\over a(t')}
.\ea
$W(x)$ is given as
\ba
\label{w}
W(x)= \frac{3 \sin(2x)}{2x^3}-\frac{3 \cos(2x)}{x^2}-\frac{3 \sin(2x)}{2x},
\ea
so that the window function $-xW'(x)$ in Eq.~(\ref{pformula}) is
%% &=&1+\frac25x^2+\dots \mbox{ for  }x \ll 1,  \\
\ba
\label{windowf}
 -xW'(x)= %\frac{d W(x)}{dx}= 
 -{3 (5 x^{2}-3)\over 2 x^{3}}\sin(2x) 
 +{3(x^{2}-3) \over x^{2}}\cos(2x), 
\ea
and the normalization is such that
\ba
\int_{0}^{\infty}\frac{dx}{x}[-x W'(x)]=1.
\ea
For the case of standard slow-roll, $G_n$ can be taken out of the integral and integration of the window function becomes unity to lead to the standard slow-roll result $n-1=-G_n$. 

We choose the fiducial model to be a flat inflaton potential which 
leads to $G_n=0$ and study 
how precisely we can constrain the deviations from $G_n=0$.  We parametrize 
these deviations as
\ba
\label{rep}
G_n(\ln \xi)=\sum_i p_i B_i(\ln \xi),
\ea
where $\{p_i\}$ form a discrete set of  parameters and $B_i(\ln \xi)$ are defined as
%\ba
% B_i(\ln \xi)=
%\cases
%      1 &  {\rm if} $\,  \ln \xi_i < \ln \xi \leq \ln \xi_{i+1} ,$ \cr
%      0 & {\rm otherwise}.
%}
%\ea
 \ba
 B_i(\ln \xi) =
\begin{cases}
1 & \text{if} \, \ln \xi_i < \ln \xi \leq \ln \xi_{i+1} , \cr
0 & \text{otherwise}.
\end{cases}
\ea
CMB observables cover less than 10 e-folds in practice corresponding to a small portion of a whole inflaton potential. We therefore, in our analysis, parametrize the inflaton 
potential across a range greater than 10 e-folds $\xi_{\rm min} \le \xi \le \xi_{\rm max}$ 
to ensure it covers the horizon crossing for observable modes.
We choose a discrete representation rather than a continuous one through 
smooth functions because, in addition to its simplicity, the discontinuities in 
$G_n(\ln \xi)$ will not show up in ${\cal P}$ due to the 
window $W(k \xi)$.  In practice, we cover a large enough range $\xi_{min}\leq \xi\leq \xi_{max}$ and choose a sufficiently fine discretization that the principal components of Sec.~\ref{principal} have converged.

\begin{figure}
    \begin{center}  
     \epsfig{file=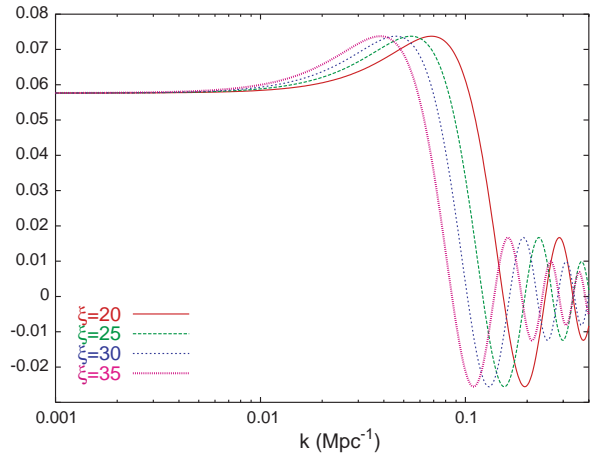, width=0.9\columnwidth}
      \caption{The response function $\partial \ln {\cal P}(k)/ \partial p_i$ as the function of $k$(Mpc$^{-1}$). The plots are given for four values of $p_i$ whose corresponding $\xi$ values are indicated.}
      \label{resp}
    \end{center}
    \end{figure}
    With this parameterization, the  power spectrum reads
\ba
\label{derive}
\ln {\cal P}=\frac{1}{f_0^2} +  
\int^{\xi_{\rm max}}_{\xi_{\rm min}}\frac{d \xi}{\xi} W(k \xi) \sum_i p_i B_i(\ln \xi),
\ea
where $1/f_0^2$ is integration constant which corresponds to the power spectrum amplitude of
the fiducial model 
\ba
{\cal P}|_{\rm fid}=\frac{1}{f_0^2}=(5.07\times 10^{-5})^2.
\ea
Here ``fid'' denotes that it is to be evaluated for the fiducial model ($i.e.$ $p_i=0$).

In  Fig. \ref{resp} we plot 
the infinitesimal response in the primordial power spectra to the 
inflation parameter $\partial \ln {\cal P}(k)/ \partial p_i$ evaluated for our fiducial model.  
Note that as might be expected the response
takes the form of a localized change in the amount of low vs high $k$ power or 
`tilt' surrounding horizon crossing $k \xi \sim 1$.

The CMB angular power spectra are given by integrals over ${\cal P}$ 
\ba
\frac{l(l+1)C_l^{XX'}}{2\pi}=\int d \ln k T^X_l(k)T^{X'}_l(k){\cal P}(k),
\ea
where $T^X_l(k)$ are the CMB transfer functions with $X$ representing the
 CMB temperature or $E$ polarization ($i.e.$ $X$ is $T$ or $E$). 
 For the numerical analysis in Sec.~\ref{principal}, we shall 
 oversample the transfer function with 3750 logarithmically spaced $k$ modes 
 from $k=10^{-4.35}-10^{-1.35}$ Mpc$^{-1}$ and 2500 $k$ modes from 
 $k=10^{-1.35}-10^{-0.35}$ Mpc$^{-1}$ (see \cite{wayne2} for details).
The infinitesimal response in the observable power spectrum to $p_i$, which 
will appear in the calculation of the Fisher matrix, is
\ba
\frac{\partial C_l^{XX'}}{\partial p_i} \Big|_{\rm fid} &=&
\frac{2 \pi}{l(l+1)}\int d \ln k T_l^X(k) T^{X'}_l(k)  \frac{1}{f^2_0} \nonumber\\
&& \times \int^{\xi_{\rm max}}_{\xi_{\rm min}}\frac{d \xi}{\xi} W(k \xi)    B_i(\ln \xi) .
\ea

We now apply Fisher matrix analysis to calculate the precision to which the $\{p_i\}$ can be
constrained to see the physical limitations for future CMB data in reconstructing 
the inflaton potential.

\section{Principal Component Analysis}

\label{principal}

The Fisher matrix analysis is useful for forecasting the uncertainties in 
parameter estimation, 
and we apply it to extract the information on an inflaton potential from $C_l$. 
Assuming a Gaussian distribution for $C_l$, the Fisher matrix is given by \cite{fis}
\ba
\label{fish}
F_{\mu\nu}=\sum^{l_{max}}_{l=2}\sum_{X,Y}
\frac{\partial C^{X}_l}{\partial p_{\mu}} 
\mathbf{Cov}^{-1}(C^{X}_{l}C^{Y}_{l})\frac{\partial C^{Y}_{l}}{\partial p_{\mu}},
\ea
where $X,Y$ represent CMB temperature, $E$ polarization or their cross correlation, 
and $\mathbf{Cov}^{-1}$ is the inverse of the covariance matrix of the power 
spectrum $C_l$. $\{p_\mu\}$ consists of the inflaton potential parameters $\{p_i\}$ 
and the cosmological parameters which are marginalized over: the dark energy density, 
equation of state parameter, reionization optical depth and the matter and baryon 
density with their corresponding fiducial values $\Omega_{DE}=0.72$, $w=-1$, $\tau=0.17$, 
$\Omega_mh^2=0.145$ and $\Omega_bh^2=0.024$ respectively.  
We additionally take the
overall amplitude of the spectrum $f_0$ as a parameter.  
This marginalization can circumvent the possible degeneracies between inflationary 
and cosmological parameters.
For quantifying the physical limitations of inflaton potential reconstruction, 
we consider ideal, cosmic variance limited observations for both temperature and 
$E$ polarization measurements
out to $l_{max}=2000$.

The inverse Fisher matrix approximates the covariance matrix 
$\mathbf{C}(\approx \mathbf{F}^{-1}$;Kramer-Rao identity) and the 
marginalized 1-$\sigma$ error $\sigma(p_\mu)$ for a given parameter $p_{\mu}$ is 
\ba
\sigma(p_\mu)=\sqrt{(\mathbf{C})_{\mu \mu}}\approx \sqrt{(\mathbf{F}^{-1})_{\mu\mu}}.
\ea
The errors in the individual parameters $p_i$ are in general large and
highly correlated to one another in part because under the general slow-roll 
approximation ${\cal P}$  picks up  contributions `around' horizon crossing 
not `at'  horizon crossing for each mode. This hinders the interpretation of 
what aspects of the potential the data will in fact constrain.
Moreover the large number of discrete parameters $\{p_{i}\}$ would make a likelihood search
with actual data unfeasible.

We therefore instead apply a principal component analysis.  Instead of $\{p_{i}\}$, we consider a new set of parameters $\{m_{i}\}$ which are linear combinations of $\{p_{i}\}$
\ba
\label{decom}
m_a=\left[\Delta\ln\xi\right]^{1/2} \sum_i S_{i a}p_{i}.
\ea
The normalization factor of $\sqrt{\Delta \ln \xi}$, which depends on the discretization of $\ln\xi$, is chosen so that the variance of
$m_a$ is independent of this spacing when it is integrated over $\ln \xi$. As long as the same convention is used for the signal and noise,
though, the normalization is arbitrary.
%The continuous representation of the above expression becomes
%\ba
%m_a\approx \int d\ln \xi \frac{S_{ia}}{\sqrt{\Delta \ln \xi}}\frac{dG(\ln \xi)}{d\ln \xi}
%\ea
The $S_{i a}$ are 
orthonormal eigenvectors of the covariance matrix of the inflation parameters 
of interest $C_{ij}$ (which is a sub-block of $C_{\mu \nu}$)
\ba
\label{eig}
\sum_{j} C_{ij}S_{j a}=\frac{1}{\Delta\ln\xi} \sigma^2_a S_{i a}.
\ea
Therefore, in our normalization the principle components $S_{ia}/\sqrt{\Delta\ln\zeta}$ 
are normalized to unit variance when integrated over $\ln \xi$
\ba
\int d \ln \xi \left[\frac{S_{ia}}{\sqrt{\Delta \ln \xi}}\right]^2
&=& \Delta \ln \xi \sum_i\left[\frac{S_{ia}}{\sqrt{\Delta \ln \xi}}\right]^2\nonumber\\
&=&1.
\ea
The
$\{m_a\}$ are orthogonal ($i.e.$ statistically independent) and their covariance matrix becomes a diagonal matrix with the elements  
\ba
\langle m_a m_b \rangle =\sigma_a^2\delta_{ab}.
\ea

We consider the parametrization of $G_n(\ln \xi)$ as given in Eq. (\ref{rep}) 
for the $\xi$ range of 
$10^{-1}<\xi / {\rm Mpc} <10^{4}$, and divide it into 200 equally spaced 
bins in $\ln \xi$ with 200 parameters $\{p_i\}$ as given by Eq. (\ref{rep}).
The principal component decomposition represented by Eq. (\ref{decom})
pinpoints the directions in the parameter space of $\{p_i\}$ with the smallest variances.

\begin{figure}
    \begin{center}  
      \epsfig{file=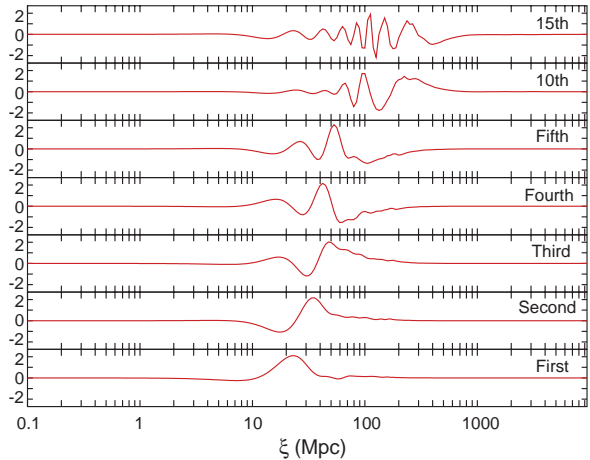, width=0.9\columnwidth}
      \caption{First five, 10th and 15th best principal components as the function 
      of the minus conformal time $\xi$.} 
%      These are the continuous versions of the modes constructed in Eq.~(\ref{decom}):
 %     $S_{ia}/\sqrt{\Delta\ln\xi}$.}
      \label{five}
    \end{center}
\end{figure}

\begin{figure}
    \begin{center}  
      \epsfig{file=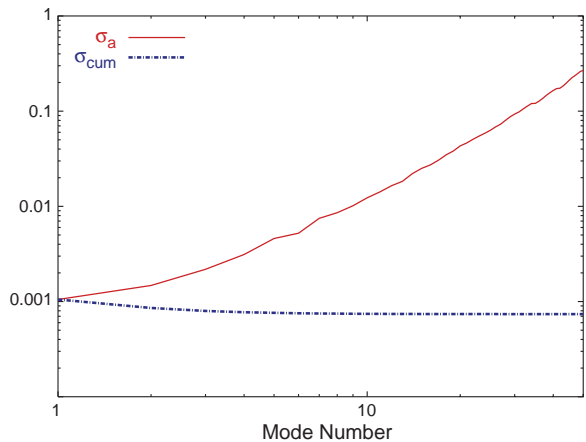, width=0.9\columnwidth}
      \caption{RMS error on each principal component as the function of mode number. Also shown is the
      square root of the cumulative variance defined in Eq.~(\ref{totalval}).}
      \label{eigval}
    \end{center}
\end{figure}

These normalized modes $S_{ia}/\sqrt{\Delta\ln\xi}$ are shown in Fig. \ref{five} which plots the five 
most tightly constrained principal components of the covariance 
matrix of the 
inflationary parameters (with the cosmological parameters marginalized over) as well as 
10th and 15th ones. 
Fig. \ref{eigval} shows the rms error $\sigma_a$ on the mode amplitudes. Since the 
principal components are all normalized to unit variance 
their joint constraining power can be estimated from the cumulative variance
\ba
\label{totalval}
\frac{1}{\sigma_{\rm cum}^2}=\sum_{a}\frac{1}{\sigma_a^2}.
\ea
Fig.~\ref{eigval} shows that
only the first handful of modes 
contribute to the cumulative variance, so we expect these to be
the only ones worth including in an analysis. In the next section, we will look at some 
specific models
and see that, practically speaking, this holds true:
all the information that can be profitably used to
compare theory with observations is contained in the first few modes. 

Note that $\xi$ is given in comoving Mpc and
its relationship to the number of e-folds from the inflationary era to the present remains
undetermined and heavily model dependent.  
%% WH is this OK?
%In Fig. \ref{five}, we would need much more information than scalar perturbations 
% for the physical interpretation of the absolute values of $\xi$ for which $C_l$ 
%can constrain an inflaton potential, since it depends on the nature of inflation 
%and the exact number of e-folds from the inflation era to the present which remain 
% heavily model dependent. 
However we can see that the first five eigenmodes are sensitive to the range of 
$10 \lesssim \xi/{\rm Mpc} \lesssim 500$ which covers $\sim \ln 50 \sim 4$ e-folds. 
This order of magnitude is consistent with our expectation from $\sim \ln l_{max} =7.6$ 
and the high cosmic variance of the low $l$ modes.

Comparing the values of $\xi$ in Fig. \ref{five} and Fig. \ref{resp} we can infer 
that the first eigenmode represents the local tilt around $k\sim 0.1$ Mpc$^{-1}$, 
which makes sense since the tilt or spectral 
index would be the best constrained aspect of ${\cal P}(k)$. This is what we would 
expect because $G_n(\ln \xi)$, which we parameterize here, represents the 
tilt as we discussed in Sec. \ref{discpa}.

\begin{figure}[h]
    \begin{center}  
      \epsfig{file=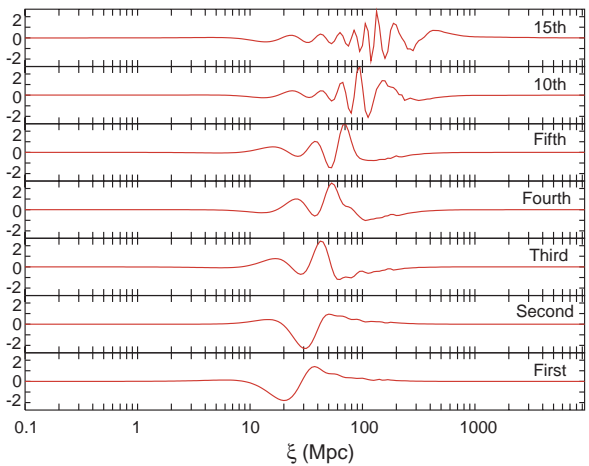, width=0.9\columnwidth}
      \caption{The best principal components when the tilt $n$ is marginalized over. 
      To be compared with Fig. \ref{five} where $n$ is not included in the marginalized 
      cosmological parameters.}
\label{secondp}    
\end{center}
\end{figure}

Indeed, if we marginalize over the tilt $n$ as well, 
the first eigenmode of Fig. \ref{five} disappears, as shown in Fig. \ref{secondp}. 
We can also infer that the second eigenmode, which has compensating contributions of opposite 
sign represents the local running ($i.e.$ the difference or
change in the local tilt).   
% Redundant
%Due to our discrete parameterization, the ``tilt" corresponds to our best constrained mode but should
%be interpreted as the local tilt around $k=0.1$. Likewise the ``running" corresponds to the local running of the tilt around $k=0.05-0.1$. This is consistent with the $k$ range of the peaks of Fig. \ref{resp} which corresponds to the $\xi$ range of the peaks for the first and second modes in Fig. \ref{resp}.   
%However, the well constrained local ``tilt" and ``running" around $k\sim 0.1$ Mpc$^{-1}$
%does not determine the properties of ${\cal P}(k)$ or $V(\phi)$ more globally or assess the
% validity of either a
%constant tilt or constant running.  This makes measuring ${\cal P}(k)$ at high
%$k$ from non CMB
%sources more interesting since in the general context, the CMB says
%nothing about
%$n(k)$ beyond the compact support region of our best eigenmodes. 
%To address the issue of the running of the running from the CMB alone, we can marginalize
%a constant running as we did the tilt and construct the best eigenmode for its detection.

%In practice the modes which are constrained depend on the data. The modes we have
%identified are for cosmic variance limited data out to $l$ of 2000. 
%Smaller data sets will
%have fewer constrained modes, and the shapes of these will be determined by the angular resolution,
%sensitivity, and sky coverage of the experiment.
%However, the formalism introduced here -- diagonalize the Fisher matrix and identify the 
%modes with smallest variance -- can easily be applied to any data set.
%Since the number of free parameters needed to describe the primordial
%spectrum using current data will certainly be less than five, this formalism is a 
%practical way to constrain
%inflation without assuming any underlying model.

Although these statements remain true for all models, understanding how constraints on 
the principal components translate into more direct statements about specific models is
more difficult.
We illustrate their utility in the next section with examples.

\section{Examples}
\label{examples}

We now apply the above formalism to two simple examples. These examples serve
two purposes: (i) they demonstrate how a generally scale-dependent $n-1$ is
mapped onto the modes defined above and (ii) they illustrate that only the top
five or so modes are likely to have signal to noise greater than one. This means
that, for the purposes of comparing with CMB observations at least, 
one can parameterize inflationary models with only five numbers, the amplitudes of the leading modes.

\begin{figure}[h]
    \begin{center}  
      \epsfig{file=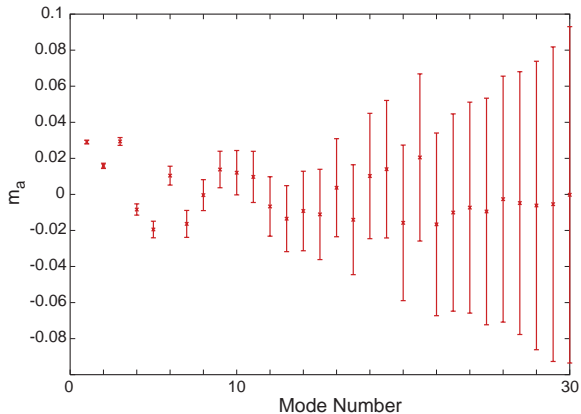, width=0.9\columnwidth}
      \caption{The first 30 eigenmodes for a potential which produces constant
      spectral index $n=0.95$. Error bars are the eigenvalues $\sigma_a$; note that only
      the first few modes have signal to noise greater than one.}
\label{sum}    
\end{center}
\end{figure}

The first example is a trivial one: suppose $n-1$ is a constant, independent of
scale. For concreteness, we choose $n=0.95$. Then $G_n=0.05$; thus all
the $p_i$'s defined in Eq.~(\ref{rep}) are equal to $0.05$. Although $G_n$ in
this model is trivial, the eigenvalues $m_a$ -- which are the convolution of
$p_i$ with the eigenvectors depicted in Figure~\ref{five} -- are not. 
Figure~\ref{sum} shows the
first thirty of these.

%\begin{figure}[h]
 %   \begin{center}  
%      \epsfig{file=005a.eps,width=0.9\columnwidth}
%      \caption{The simple case of constant $n=0.95$ gives no change of $dG/d\ln \xi$ as the function 
%      of $\xi$.}
%\label{pot}    
%\end{center}
%\end{figure}

 This example shows that every inflation model can be
expressed as a set of ${m_a}$. Because each $m_a$ is independent, 
we can obtain the total signal to noise by calculating 
$\sqrt{\sum_i m_i^2/\sigma_i^2}$. We found, for this example of $n-1=-0.05$, 
$\sum_{i=1}^{2000} m_i^2/\sigma_i^2=1102$ which gives us the estimate of 
$n-1\approx -0.05(1 \pm 1/\sqrt{1102})=-0.05 \pm 0.0015$. This can be compared
with the variance of $n$ obtained directly from the Fisher matrix (where $n$ is taken as
the only inflation parameter).
In this case, we find $\sigma(n)=0.0015$, in perfect agreement with the
more general approach.
We also found that including only the first five modes
in the calculation of total signal to noise gives $\sum_{i=1}^{5} m_i^2/\sigma_i^2=1088$. 
Hence, as long as
the top five modes are retained, we retain essentially all the information about 
$n$. If we knew that $n$ were constant of course, it would be better to
use the constant $n$ as the variable to be compared with data. 

For more general cases including the case of a variable $V''$, the proper way to fit the data would be to allow the best constrained 
optimal eigenvalues, more specifically five optimal ones, to be free parameters. Our next example demonstrates that this does in fact work. 

\begin{figure}[h]
    \begin{center}  
      \epsfig{file=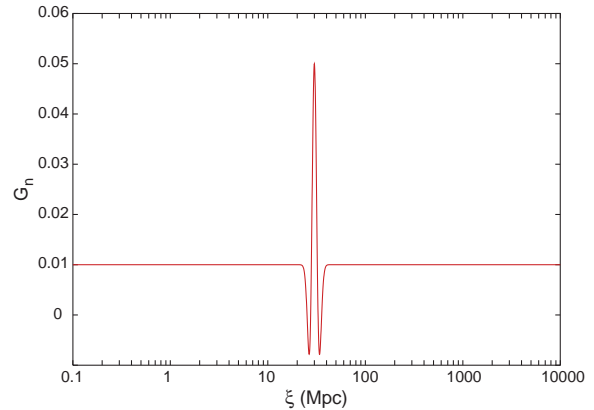,width=0.9\columnwidth}
      \caption{The change of $G_n$ as the function of $\xi$ for a potential 
      with a bump given by Eq. (\ref{eq9}) for the case of $\lambda=0.1, c=10^{-6}, \nu=100$ and $\xi_0=30$.}
\label{pot}    
\end{center}
\end{figure}

Consider a perturbation at $\phi=\phi_0$ in an inflaton potential 
represented by a smooth bump 
\ba
\label{eq9}
V(\phi)=V_0e^{\lambda(\phi-\phi_0)}[1+ c e^{-\nu^2(\phi-\phi_0)^2} ] 
\ea
where $ |\lambda| \ll 1,|c|\ll 1, |\nu| \gg 1$.
We now have
\ba
\label{eq11}
\frac{d \phi}{d \ln (aH)}&=&-\frac{V'}{V}\left[1+{\cal O}\left( \frac {{V'}^2}{V^2}\right)\right]\nonumber  \\
&=&-\lambda~~~ \mbox{to the leading order}
\ea
so that $\phi \simeq \lambda \ln \xi$ (an integration factor is scaled into $\xi$ such that 
$\phi_0=\lambda \ln \xi_0$.)
Eqs. (\ref{eq9}) and (\ref{eq11}) can be substituted into Eq.~(\ref{para})
to obtain $G_n$ as a function of $\xi$. This is shown in
Fig.~\ref{pot} for the parameters $\lambda=0.1, c=10^{-6}, \nu=100$ and $\xi_0=30$.

\begin{figure}[h]
    \begin{center}  
      \epsfig{file=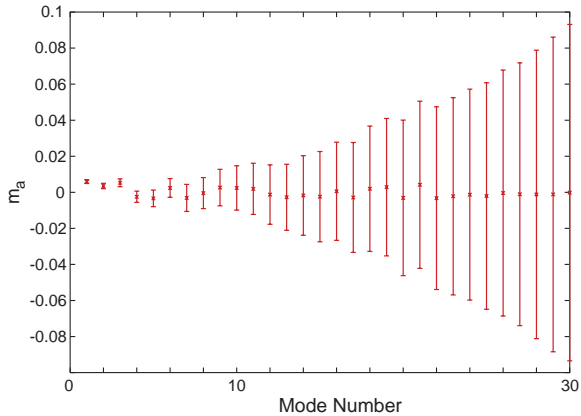, width=0.9\columnwidth}
      \caption{The corresponding first 30 eigenmodes for the potential with a bump  given by Eq. (\ref{eq9}) for the case of $\lambda=0.1, c=10^{-6}, \nu=100$ and $\xi_0=30$.}
\label{sum1}    
\end{center}
\end{figure}
 
 Convolving the $\{p_i\}$ shown in  Fig.~\ref{pot}  with the
 eigenmodes in Fig.~\ref{five}, we obtain the $\{m_a\}$ for this model; 
 these are shown in Fig.~\ref{sum1}. 
 We see that only the first few modes have signal to noise bigger than 
 unity. They therefore are all that need to be retained
 to capture the information necessary to distinguish 
this non-trivial potential from a perfectly flat potential. This is 
reasonable because, in this example, the deviation of $G_n(\ln \xi)$ from 
zero occurs around the range of $\xi$ to which only the well-constrained modes are 
sensitive.

\section{Conclusion and Discussion}
\label{conc}

We applied a principal component analysis to the covariance matrix to investigate the 
precision with which CMB observables can ultimately constrain the inflaton potential,
accounting for the uncertainties in
cosmological parameters.
 We parameterized the inflaton potential using 200 parameters in the framework of the general 
 slow-roll approximation without specifying any particular inflationary scenario. We showed that considering five principal components covering about four e-folds of conformal time will be sufficient to obtain almost all of the information on the features of the inflaton potential when the deviation from the flat potential is within $|G_n(\ln \xi)| \lesssim 0.05$ in the scales CMB observables are sensitive to. We found the first principal component represents the 
 local tilt and the second one the local running in the region around $k=0.05-0.1 $ 
 Mpc$^{-1}$ where
 cosmic variance 
 limited CMB observations out 
 to $l=2000$ are most constraining. 

We, however, should keep in our minds that the well constrained local ``tilt" and ``running" around $k\sim 0.1$ Mpc$^{-1}$
does not determine the properties of ${\cal P}(k)$ or $V(\phi)$ more globally or assess the
 validity of either a
constant tilt or constant running.  This makes measuring ${\cal P}(k)$ at high
$k$ from non CMB
sources more interesting since in the general context, the CMB says
nothing about
$n(k)$ beyond the compact support region of our best eigenmodes. Our formalism is general and can be easily extended to multiple data sets. For example, with small scale power spectrum
information it may be interesting to marginalize a constant running as we did the tilt and construct the best eigenmode for the running of the running which can become rather important well beyond the few e-folds of the CMB range in the near future.

 The principal component analysis helps clarify the
 particular aspects of the inflation potential that are well constrained. It is also a useful 
 technique to find the optimal way of extracting information on the
inflaton potential in a model independent manner in the analysis of actual data, say, via the 
Markov Chain Monte Carlo method once the polarization and higher angular resolution data
becomes available in the future. 
\\
\\
\\
%\subsection*{Acknowledgments}    
This work was supported by DOE, by NASA grant NAG5-10842, 
by the Packard Foundation, and by ARCSEC funded by Korea Science and Engineering Foundation and the Korean Ministry of Science, 
the Korea Research Foundation grant KRF PBRG 2002-070-C00022 and Brain Korea 21.

\section*{Appendix: General Slow-roll Approximation}

\begin{figure} 
 \centering 
      \epsfig{file=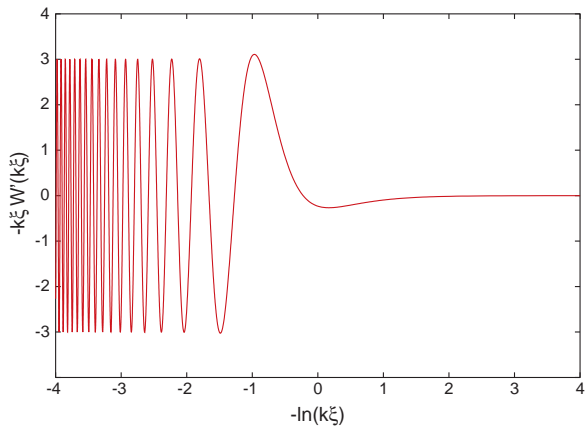, width=0.9\columnwidth}
 \caption{The window function for the general slow-roll formula $-k\xi W'(k\xi)$ as the function of $-\ln (k\xi)$. Large $k\xi$ corresponds to earlier times, when the mode of interest
 is within the horizon, so time flows from left to right.} 
 \label{window} 
\end{figure}

Under the general slow-roll approximation, the power spectrum for a single field 
inflation model is given as
\ba
\label{pformula}
\ln {\cal P}(\ln k) &=&\int^{\infty}_{0}\frac{d \xi}{\xi}\left[-k \xi W'(k \xi)    \right]\left[\ln \frac{1}{f^2} +\frac{2}{3} \frac{f'}{f}    \right] \nonumber\\
&\equiv& \int^{\infty}_{0}\frac{d \xi}{\xi}\left[-k \xi W'(k \xi)    \right] G(\ln \xi) ,
\ea
The window function is defined in Eq. (\ref{windowf}).
The dependence on the dynamics of the inflaton field is encoded in 
\ba
f(\ln \xi)\equiv \frac{2 \pi a \xi \dot{\phi}}{H} ,
\ea 
where $H$ is the Hubble constant and  $\dot{\phi}=d \phi/d t$ is the time derivative of the inflaton field.  
Here and below primes denote derivatives with respect to the argument, e.g.
$f'=df/d \ln \xi$. $\xi=(aH)^{-1}[1+{\cal O}(\dot{H}/H^2)]$, so, to the leading order, $f=2 \pi \dot{\phi}/H^2$ 
which is just an inverse of the familiar comoving curvature perturbation. The amplitude of $f'/f$ is small in the general slow-roll approximation as well as in the standard slow-roll approximation. $f'/f$ can, however, vary rapidly in the general slow-roll approximation in contrast to the case of the standard slow-roll approximation. 
As we can see from the integral over $d \ln \xi$ in Eq. (\ref{pformula}), the power
spectrum receives contributions from a range of $\xi$ for a given mode $k$, not only the
one moment of horizon crossing as in the standard slow-roll formula.
This is because a given Fourier mode is affected by the possibly prolonged 
inflationary dynamics while it is leaving the horizon, not just at one instant; these details are specified 
by the properties of the window function $-k\xi W'(k\xi)$ which is shown in Fig. \ref{window}.
The mode oscillates rapidly inside the horizon (when $k\xi \gg 1$), and one would need a
sudden event to affect it. The window function starts to vanish for $ k\xi < 1$ once 
the mode leaves the horizon because it freezes out and becomes hard to affect. 
For a slowly varying $f'$ as in the case of the standard slow-roll, the window
function is not resolved and just appears as a delta function. Then, Eq.~(\ref{pformula})
reduces to ${\cal P}=1/f^2=(H^2/2\pi\dot{\phi})^2$, the standard result~\cite{ll,mc}.
$G_n$ in the body of the paper is nothing but the derivative of the integrand in the general slow-roll 
formula Eq. (\ref{pformula}) as follows\footnote{We are here assuming a smoothly varying $V(\phi)$, see also \cite{ewangeneral}.}
\ba
\label{para}
\frac{d}{d \ln \xi}G(\ln \xi)&\equiv& \frac{d}{d \ln \xi} \left(  \ln \frac{1}{f^2} +\frac{2}{3} \frac{f'}{f}     \right) \nonumber\\&=&
\frac{2}{3}\left(\frac{f''}{f}-3\frac{f'}{f} -\frac{{f'}^2}{f^2}\right) \nonumber \\
&=& \frac{2}{3}\left(\frac{f''}{f}-3\frac{f'}{f} \right)~~ \mbox{ to the leading order}
\nonumber\\
&=&
 3\left(\frac{V'}{V}\right)^2-2 \frac{V''}{V},
\ea
which is true even if $V''$ changes with $\xi$ \cite{ewangeneral}. 
\begin{figure} 
 \centering 
      \epsfig{file=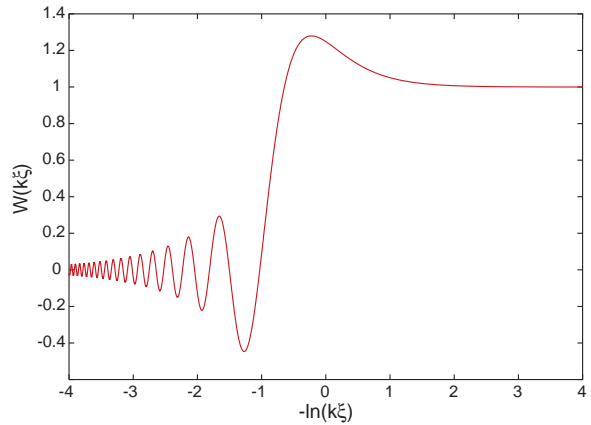, width=0.9\columnwidth}
 \caption{The window function $W(k\xi)$ as the function of $-\ln (k\xi)$} 
 \label{wplot} 
\end{figure}
Let us now derive Eq. (\ref{derive}). Covering a large enough $\xi$ range $\xi_{min}\leq \xi \leq \xi_{max}$ to ensure it covers the horizon crossing for observable modes of our interest, the right hand side of Eq. (\ref{pformula}) can be reduced to
%\ba
% \ln {\cal P}(\ln k)= &&  \int^{\infty}_{\xi_{\rm max}} \frac{d \xi}{\xi} \left[-k \xi W'(k \xi)    \right]G(\ln \xi) \nonumber \\
% && +  \int^{\xi_{\rm max}}_{\xi_{\rm min}}\frac{d \xi}{\xi} \left[-k \xi W'(k \xi)    \right]G(\ln \xi)\nonumber\\
%&&+ \int^{\xi_{\rm min}}_{0}\frac{d \xi}{\xi} \left[-k \xi W'(k \xi)    \right]G(\ln \xi).
%\ea
%Because 
%$W'(k\xi)\rightarrow 0$ for sufficiently large  $\xi_{\rm max} \gg k^{-1}$ as well
%as for sufficiently small $\xi_{\rm min} \ll k^{-1}$, 
%we can ignore the negligible contributions from the large and small $\xi$ in the above to obtain
\ba
\label{ap}
\ln {\cal P} =  \int^{\xi_{\rm max}}_{\xi_{\rm min}} \frac{d \xi}{\xi} \left[-k \xi W'(k \xi)    \right]G(\ln \xi),
\ea
with negligible contributions from $\xi$ outside the limits of integration which can be justified from $W'(k\xi)\rightarrow 0$ for the large enough $\xi_{max}$ and the small enough $\xi_{\rm min}$.
Integrating by parts, 
\ba
\ln {\cal P}&=& 
- [W(k \xi)G(\ln \xi)]^{\xi_{\rm max}}_{\xi_{\rm min}} \nonumber\\ &&+
 \int^{\xi_{\rm max}}_{\xi_{\rm min}} \frac{d \xi}{\xi} W(k \xi)\frac{d G(\ln \xi)}{d \ln \xi} .
\ea
$W(k\xi)$ as the function of $-\ln (k\xi)$ is shown in Fig. \ref{wplot}. $W\rightarrow 0$ for the large enough $\xi_{max}$ and $W\rightarrow 1$ for the small enough $\xi_{min}$ so that
\ba
\ln {\cal P}= 
G(\ln \xi_{\rm min})+
 \int^{\xi_{\rm max}}_{\xi_{\rm min}} \frac{d \xi}{\xi} W(k \xi)\frac{d G(\ln \xi)}{d \ln \xi} .
\ea
where the integration constant $G(\ln \xi_{\rm min})$ gives the normalization amplitude of ${\cal P}$.

\newpage

%%%%%%%%%%%%%%%%%%%%%%%%%%%%%%%%%%%%%%%

\end{document}